\documentclass[prl,twocolumn,superscriptaddress]{revtex4-1}
\usepackage{graphicx}
\usepackage{dcolumn}
\usepackage{amsmath, amssymb}
\usepackage{siunitx}
\usepackage{bm}
\usepackage{textcomp}
\newcommand{\tsim}{{\raise.19ex\hbox{$\scriptstyle\sim$}}}

\begin{document}

\title{Superlattice-induced insulating states and valley-protected orbits in twisted bilayer graphene}
\author{Y. Cao}
\author{J. Y. Luo}
\author{V. Fatemi}
\affiliation{Department of Physics, Massachusetts Institute of Technology, Cambridge, Massachusetts 02139, USA}
\author{S. Fang}
\affiliation{Department of Physics, Harvard University, Cambridge, Massachusetts 02138, USA}
\author{J. D. Sanchez-Yamagishi}
\affiliation{Department of Physics, Harvard University, Cambridge, Massachusetts 02138, USA}
\author{K. Watanabe}
\author{T. Taniguchi}
\affiliation{National Institute for Materials Science, Namiki 1-1, Tsukuba, Ibaraki 305-0044, Japan}
\author{E. Kaxiras}
\affiliation{Department of Physics, Harvard University, Cambridge, Massachusetts 02138, USA}
\affiliation{John A. Paulson School of Engineering and Applied Sciences, Harvard University, Cambridge, Massachusetts 02138, USA}
\author{P. Jarillo-Herrero}
\email{pjarillo@mit.edu}
\affiliation{Department of Physics, Massachusetts Institute of Technology, Cambridge, Massachusetts 02139, USA}

\date{\today}
\begin{abstract}
Twisted bilayer graphene (TwBLG) is one of the simplest van der Waals heterostructures, yet it yields a complex electronic system with intricate interplay between moir\'{e} physics and interlayer hybridization effects. We report on electronic transport measurements of high mobility small angle TwBLG devices showing clear evidence for insulating states at the superlattice band edges, with thermal activation gaps several times larger than theoretically predicted. Moreover, Shubnikov-de Haas oscillations and tight binding calculations reveal that the band structure consists of two intersecting Fermi contours whose crossing points are effectively unhybridized. We attribute this to exponentially suppressed interlayer hopping amplitudes for momentum transfers larger than the moir\'{e} wavevector.

\end{abstract} 

\maketitle
The plethora of available two dimensional materials has led to great interest in investigating novel quantum phenomena that can originate from assembling them into van der Waals heterostructures \cite{geim2013}. One of the simplest such heterostructures is twisted bilayer graphene (TwBLG), consisting of two sheets of monolayer graphene stacked on top of each other with a relative twist angle. Despite the material simplicity, an intricate interplay between moir\'{e} physics and interlayer hybridization effects exists in TwBLG --- one striking consequence is that the heterostructure can host an insulating state even though it comprises two sheets of high quality conductors. The intrinsic band gap is due to interlayer hybridization; this is in contrast to the graphene/hexagonal boron nitride (h-BN) moir\'{e} heterostructure where the band gap at charge neutrality \cite{hunt2013,woods2014} arises from other mechanisms such as sublattice symmetry breaking, strain effects, and many-body interactions \cite{jung2015,sanjose2014,song2013}).

Due to the different orientation of the two graphene lattices in TwBLG, a periodic modulating potential related to the resultant superlattice moir\'{e} pattern emerges. Furthermore, the bands in both graphene layers can readily hybridize and exhibit strong interlayer coupling \cite{santos2007,mele2010,mele2011,bistritzer2011}. The extent of the hybridization depends critically on the relative twist angle $\theta$. For $\theta>\SI{3}{\degree}$, the Dirac cones of the two layers are separated far apart in momentum space, and hybridization occurs at high energies and densities \cite{ohta2012,havener2014} which are typically inaccessible in transport experiments. On the other hand, for small $\theta$, hybridization occurs at low energies between nearby $K$ points of opposite layers, leading to a drastically reduced Fermi velocity which has been confirmed by scanning tunneling microscopy experiments \cite{li2010,luican2011,yan2012,brihuega2012}.

\begin{figure}[tb]
	\includegraphics[width=\columnwidth]{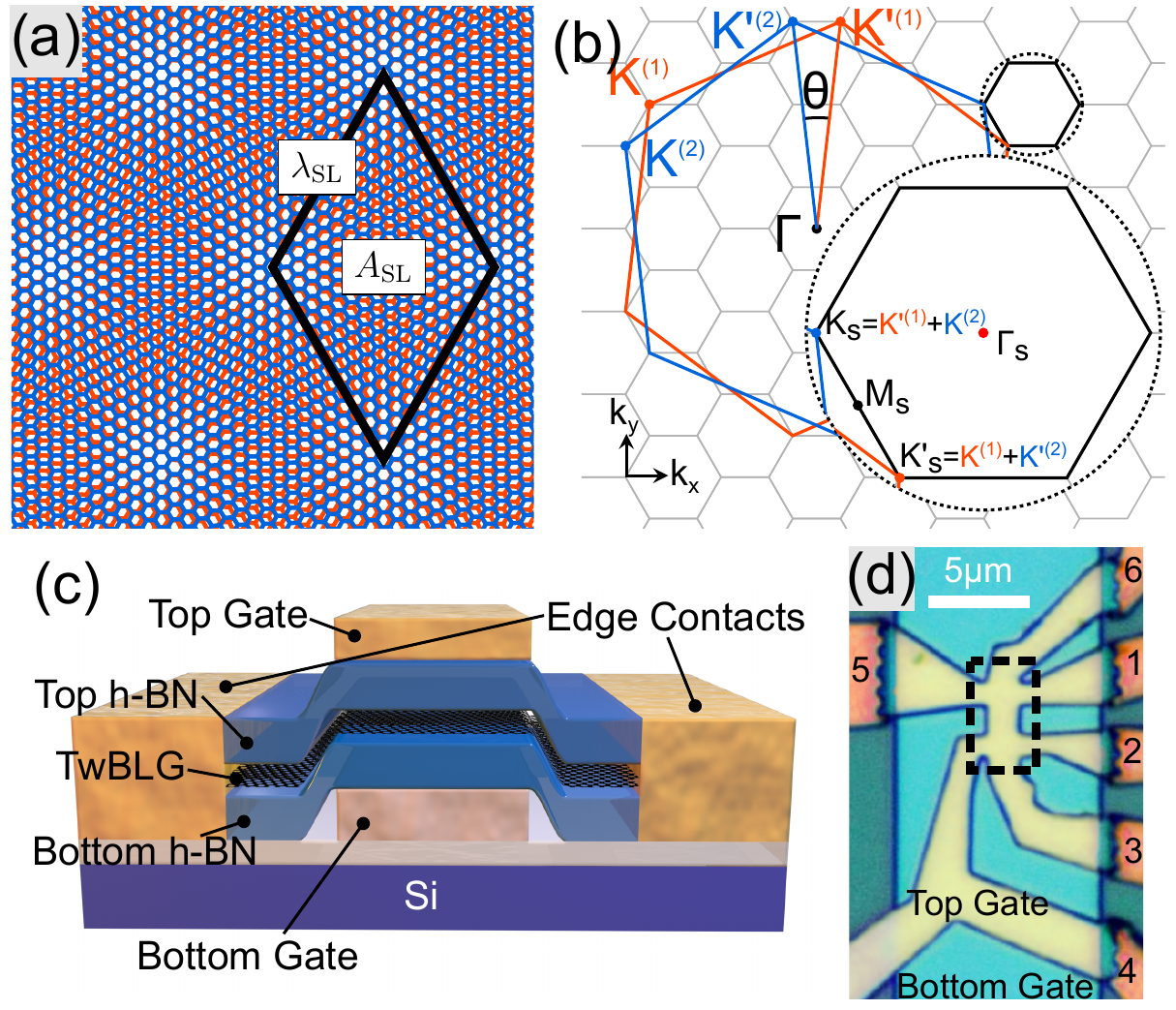}
	\caption{\label{fig:fig1} (color online). (a) Schematic of TwBLG and its superlattice unit cell. $\lambda_{\mathrm{SL}}=\frac{a}{2\sin(\theta/2)}$ ($a$ is the lattice constant of graphene) is the moir\'{e} period and $A_{\mathrm{SL}}=\frac{\sqrt{3}}{2}\lambda_{\mathrm{SL}}^2$ is the unit cell area. (b) The orange and blue hexagons denote the original Brillouin zones of graphene layer 1 and 2 respectively. In $k$-space, the band structure is folded into the MBZ which is defined by the mismatch between the hexagonal Brillouin zones of the two honeycomb lattices. (c) Illustration of the cross section of our device. (d) Optical image of $\theta\!\approx\!\SI{1.8}{\degree}$ device S1. The hall bar in the dashed rectangular region is completely free of bubbles and ridges. }
\end{figure}

\begin{figure}[tb]
	\includegraphics[width=\columnwidth]{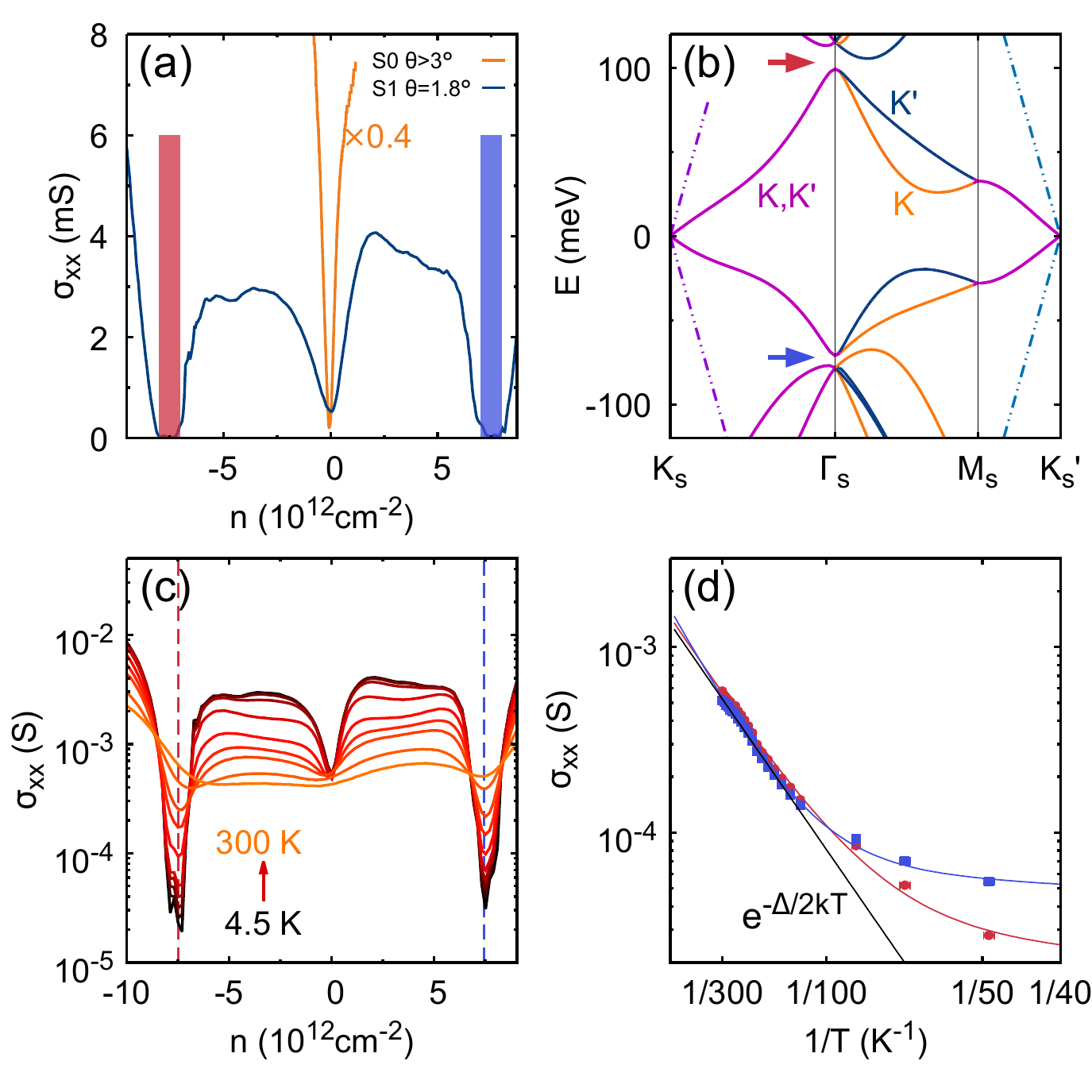}
	\caption{\label{fig:fig2} (color online). (a) Comparison of the conductivity of a large angle TwBLG device S0 and a small angle device S1. The vertical bars around $n=\SI{+-7.5e12}{\per\centi\meter\squared}$ indicate the insulating states in device S1. (b) Tight-binding band structure of TwBLG with $\theta=\SI{1.8}{\degree}$. Dashed lines denote the monolayer graphene dispersion with Fermi velocity $v_F=\SI{1e6}{\meter\per\second}$. The color of the bands denotes the valley polarization: $K$ (orange), $K^\prime$ (navy blue), and valley-degenerate (purple). The arrows indicate the direct band gaps at $\Gamma_s$. (c) Temperature dependent conductivity of device S1. (d) Arrhenius plot of the conductivity of the insulating states [indicated by dashed lines in (c)]. Blue and red denote the electron and hole side insulating states respectively. Thermal activation gaps of $\tsim~$\SI{50}{\milli\electronvolt} and $\tsim~$\SI{60}{\milli\electronvolt} are estimated from the slope for the electron-side and hole-side insulating states respectively. }
\end{figure}

In a moir\'{e} superlattice [Fig.~\ref{fig:fig1}(a)], the band structure must be reconsidered in a mini Brillouin zone (MBZ) that corresponds to the superlattice unit cell, as shown in Fig.~\ref{fig:fig1}(b). At low twist angles, theory suggests that the interlayer interaction significantly distorts the band structure of TwBLG, such that the system can no longer be described by two weakly coupled Dirac cones at low energies (which is valid for large angle TwBLG) \cite{morell2010,moon2012}. In particular, various calculations predict that a single-particle gap can be opened at the $\Gamma_s$ point of the MBZ in a specific range of twist angles when the lowest energy superlattice bands are filled [Fig.~\ref{fig:fig2}(b)] \cite{moon2012,fang2016}. This can be understood to arise from the strong interlayer coupling in small angle TwBLG, which allows for substantial interlayer Bragg reflections off the superlattice potential. On the other hand, a long-range periodic potential in itself is insufficient to open a gap at the superlattice points in a graphene/h-BN heterostructure, as there is no low-energy state in h-BN that can couple to the graphene bands \cite{wallbank2013}.  Despite these theoretical predictions for TwBLG, no experimental evidence to date directly points to the existence of global energy gaps when the superlattice bands are completely filled \cite{lee2011,schmidt2014}.

In this Letter, we report observations of insulating states at the superlattice band edges in small angle TwBLG via transport measurements, where we measure thermal activation gaps of \SI{50}{\milli\electronvolt} and \SI{60}{\milli\electronvolt} on the electron and hole sides respectively. Additionally, in the quantum Hall regime, the eight-fold degeneracy of the Dirac points transitions to a four-fold degeneracy near the superlattice band edge. Finally, by comparing Shubnikov-de Haas oscillations with a tight-binding model, we deduce that the band structure consists of two intersecting Fermi contours whose crossing points are essentially not hybridized due to the exponentially suppressed hopping amplitudes for momentum transfers much larger than the moir\'{e} wavevector. \cite{shallcross2013,supplementary,bistritzer2011}.

We fabricated fully-encapsulated TwBLG devices with $\theta<\SI{2}{\degree}$ using a modified dry-transfer method \cite{wang2013}. The samples are dual-gated for independent control of the total charge density and interlayer potential difference \cite{supplementary}. A local metallic bottom gate is used to screen the charge impurities present in the silicon oxide substrate, and one-dimensional edge contacts are used to contact the TwBLG \cite{wang2013}.  A `tear-and-stack' technique was also developed to enable sub-degree control of the twist angle \cite{supplementary,kim2016}. We used an \emph{ab initio} tight-binding model for the calculation of band structures and related quantities \cite{supplementary,fang2016}. 

Our samples show Hall mobilities exceeding \tsim\SI{20000}{\centi\meter\squared\per\volt\per\second} at $T=\SI{4}{\kelvin}$. Fig.~\ref{fig:fig2}(a) shows the conductivity of two TwBLG samples: sample S1 with a low twist angle (we focus on device S1 in this Letter, but data on other small angle TwBLG devices with similar behaviors are also presented in the supplement \cite{supplementary}) and sample S0 with a large twist angle ($>\SI{3}{\degree}$). In both samples, the conductivity minimum centered at zero density corresponds to the degenerate Dirac points in both layers of graphene. However, for the small angle sample S1, we observe two insulating states occurring at total carrier densities of $n\!\approx\!\pm\SI{7.5e12}{\per\centi\meter\squared}$, which are symmetric on both sides of the charge neutrality point.   

We attribute these insulating states to the gaps occurring at the $\Gamma_s$ point of the MBZ when the lowest-energy superlattice bands are fully filled. The \emph{ab initio} tight-binding calculation of the commensurate $\theta=\SI{1.8}{\degree}$ TwBLG is shown in Fig.~\ref{fig:fig2}(b). The low-energy bands of TwBLG retain the valley polarizations of its constituent graphene layers, \emph{i.e.} valley continues to be a valid quantum label for these bands. The bands colored in orange correspond to $K$-valleys, while the blue bands correspond to $K^\prime$-valleys. Although the $K$ valley of one graphene layer and the $K^\prime$ valley of the other layer occupy the same $k$-points in the MBZ along the $\Gamma_s$---$K_s$ line (purple lines), their hybridization is suppressed because of the large momentum mismatch in the original graphene Brillouin zone, as explained later in this Letter. Therefore, valley still provides a 2-fold degeneracy even far away from the Dirac point, and the total density required to fill up to the insulating gaps is equal to $4$ times the MBZ area: 2 from the valley quantum number and 2 from spin. From the deduced density $n=\pm\SI{7.5e12}{\per\centi\meter\squared}$ at the center of the insulating states, we derive the unit cell area of the superlattice to be $A_\textrm{SL}=4/n=\SI{53.3}{\nano\meter\squared}$, with a corresponding twist angle of $\theta=\SI{1.8}{\degree}$. This agrees well with our target value of $\theta=\SI[separate-uncertainty = true]{2+-0.5}{\degree}$. 

To study thermally activated transport of the insulating states, we measured the temperature dependence of the conductivity of sample S1 [Fig.~\ref{fig:fig2}(c-d)].  The insulating states' conductivities drop by more than an order of magnitude from \SI{300}{\kelvin} to \SI{50}{\kelvin}, and start to saturate below \tsim\SI{50}{\kelvin}. An Arrhenius-like behavior is evident at higher temperatures. From the slope in the Arrhenius plot between \SI{100}{\kelvin} and \SI{300}{\kelvin}, we estimate the thermal activation gaps to be $\sim\SI{50}{\milli\electronvolt}$ and $\sim\SI{60}{\milli\electronvolt}$ for the electron-side and hole-side insulating states respectively. The deviation from Arrhenius-like behavior at low temperatures may be attributed to a variable-range hopping mechanism \cite{mott1968,supplementary}.

Our experimental observations differ from theoretical predictions in two important ways. First, we observe an insulating state on the hole side despite calculations not predicting a global transport gap at negative energies, as shown in Fig.~\ref{fig:fig2}(b). Second, although calculations predict a transport gap at the $\Gamma_s$ point of the MBZ at positive energies \cite{moon2012,fang2016}, the measured activation gap is much larger than expected. Contributing factors may include an underestimation of the interlayer interaction strength in these calculations, but these are unlikely to account for most of the difference. Physical effects of lattice strain, as recently proposed to explain the energy gap in monolayer graphene/h-BN structures, may also play a significant role \cite{jung2015,sanjose2014}. A third possibility is an excitonic instability, as reported for Bernal bilayer graphene \cite{jerome1967,nandkishore2009,bao2012}. The small single-particle gap and the 2D nature of the system make it possible for the excitonic binding energy to be the larger energy scale.

\begin{figure}[tb]
	\includegraphics[width=\columnwidth]{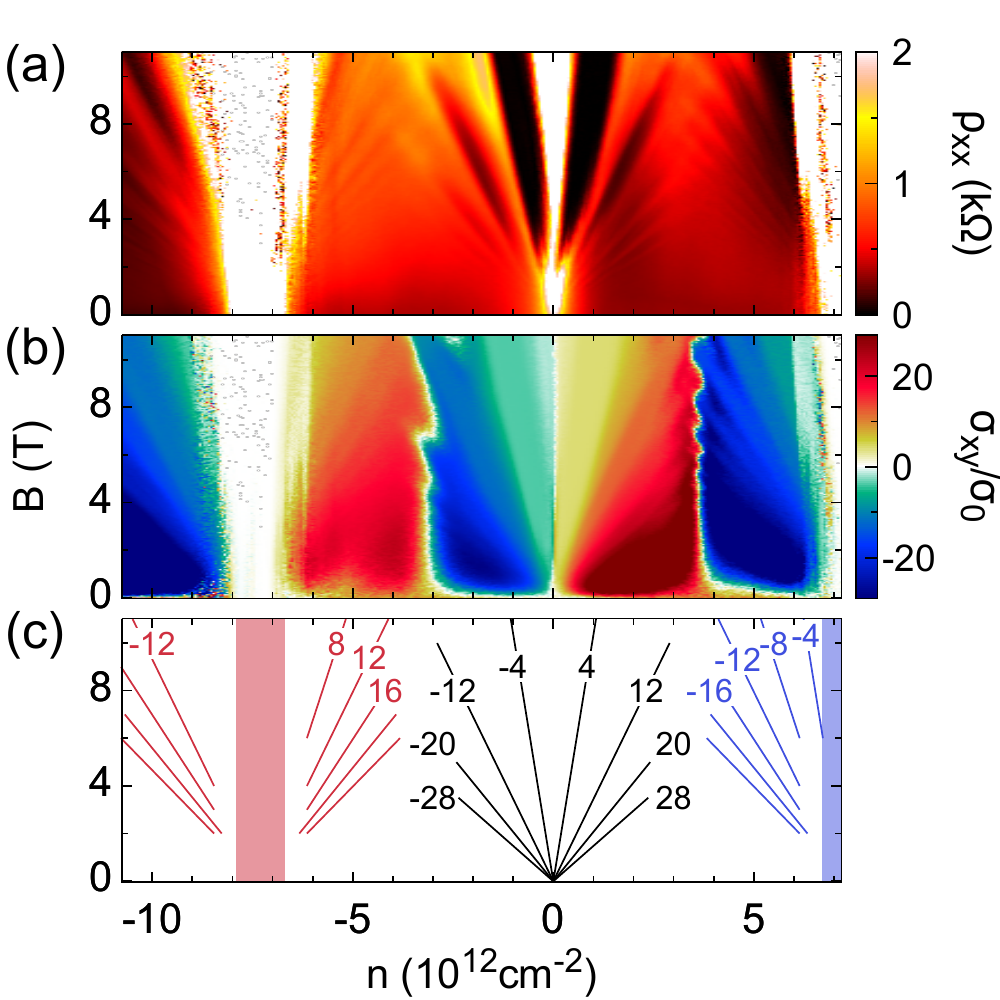}
	\caption{\label{fig:fig3}(color online). (a) Longitudinal resistivity and (b) Hall conductivity versus the total density and the magnetic field.  Measurements are taken at $T=\SI{40}{\milli\kelvin}$. (c) Reconstructed Landau level structure from the plateau values. The central Landau fan emanating from the Dirac point at zero density has an 8-fold degenerate half-integer quantum Hall sequence, while the Landau fans originating from the superlattice gaps have a 4-fold degenerate massive parabolic quantum Hall sequence.   }
\end{figure}

Next, we apply a perpendicular magnetic field to the TwBLG sample. Fig.~\ref{fig:fig3} shows the longitudinal resistivity, $\rho_{xx}$, and the Hall conductivity, $\sigma_{xy}$, as a function of the total density $n$ and the magnetic field $B$. In a magnetic field, the Hall conductivity quantizes according to $\sigma_{xy}=\nu e^2/h$, with the filling factor $\nu=n \phi_0/B$, where $\phi_0=h/e$ is the flux quantum. The central Landau fan that originates from the Dirac cone near zero density generates filling factors of $\nu=\pm4, \pm12, \pm20$, \ldots. This sequence is double that of the monolayer graphene  quantum Hall sequence of $\nu=\pm2, \pm6, \pm10, \ldots$, indicating that at low energies a massless Dirac dispersion is retained despite the strong interlayer hybridization \cite{jdsy2012,degail2011}. 

\begin{figure}[!htb]
	\includegraphics[width=\columnwidth]{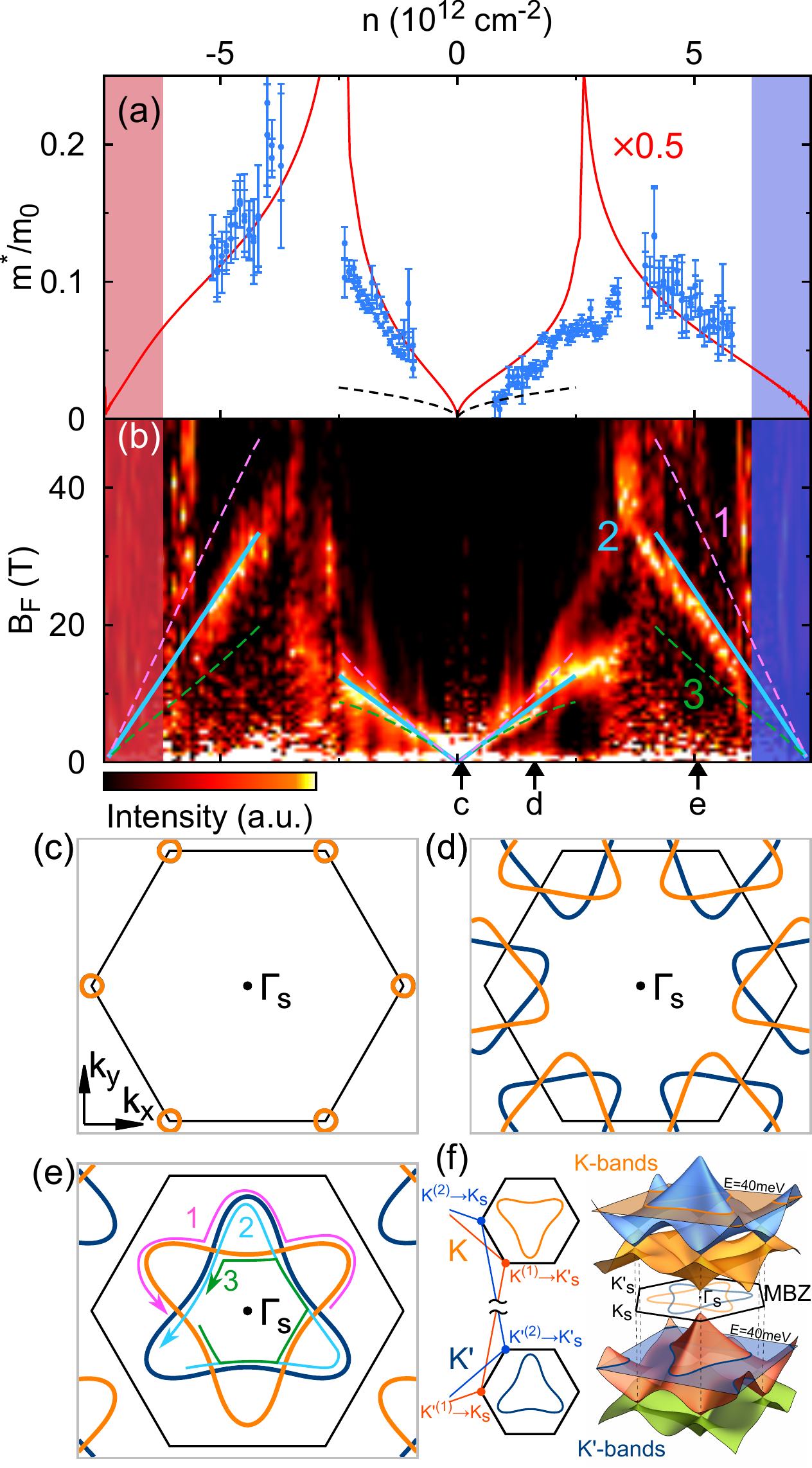}
	\caption{\label{fig:fig4} (color online). (a) Cyclotron masses and (b) oscillation frequencies extracted from SdH measurements. The red curve is the numerically calculated cyclotron mass (normalized by a factor of 0.5) and the black dashed curve is the effective mass if the interlayer interaction is ignored. Lines colored pink, blue and green denote the expected slope for the outer star orbit, triangular orbits and inner hexagon orbit shown in (d) and (e). (c-e) Fermi contours at densities shown as arrows positioned below the density axis in (b). Orange orbits are $K$-polarized, and blue orbits are $K^\prime$-polarized. (f) 3-D Illustration of the low-energy band structure. The two sets of bands are valley-polarized in the original $K$, $K^\prime$ valleys of the constituent layers. For example, the $K$ sub-bands result from the hybridization of $K^{(1)}$ and $K^{(2)}$ Dirac cones. The same applies for the $K^\prime$ sub-bands.}
\end{figure}

However, the Landau fans originating from the insulating states differ markedly from the massless Dirac nature of the central Landau fan. As shown in Fig.~\ref{fig:fig3}(c), the Landau level sequence near the insulating states is $\nu=0, \pm4, \pm8, \pm12, \pm16, \ldots$, indicating a non-Dirac massive band \cite{kim2016_2}. The 4-fold degeneracy of this sequence is attributed to the spin degeneracy and the Fermi contour degeneracy from the valley quantum number near the $\Gamma_s$ point. The lack of a Berry phase on the other hand indicates a parabolic band edge at the insulating states \cite{moon2012}. Additionally, we observe a sign change of $\sigma_{xy}$ at $n\!\approx\!+ (-)\SI{3e12}{\per\centi\meter\squared}$, indicating a transition of massless Dirac electron-like (hole-like) carriers to massive hole-like (electron-like) carriers.   

We further investigate this transition by examining the density of states $D(E)$ in TwBLG through Shubnikov-de Haas (SdH) oscillations. By fitting the temperature-dependence of the SdH oscillation amplitude to the Lifshitz-Kosevich formula, we can obtain the cyclotron mass $m^*$ at the Fermi energy, which for a two-dimensional system is proportional to the density of states per Fermi pocket at the Fermi energy, \emph{i.e.} $m^*=\frac{h^2}{2\pi}D(E)/N$, where $N$ is the degeneracy. The blue data points in Fig.~\ref{fig:fig4}(a) are the extracted cyclotron masses as a function of total density, while the red curve is the numerically calculated cyclotron mass multiplied by a normalization factor. For TwBLG, $m^*$ is expected to peak at the van Hove singularities \cite{kim2016_2} and to approach zero at both the Dirac point and the superlattice gaps. This is consistent with our observation that the slope of $m^*$ vs. density changes sign, in correspondence to the sign of the charge carrier extracted from Hall measurements.

Additionally, we find that near the Dirac points, $m^*$ is about 2.5 times larger than that of monolayer graphene, indicating a similar reduction in the Fermi velocity as observed in other studies \cite{li2010,luican2011,yan2012,brihuega2012}. The experimental and theoretically calculated $m^*$ agree well up to a uniform scaling factor for all densities, which may be attributed to underestimation of the band width in the \emph{ab initio} calculations \cite{fang2015} or to corrections to the $m^*$ term in the Lifshitz-Kosevich formula for 2D systems \cite{martin2003}. 

Further information about the band structure is obtained from analysis of SdH oscillation frequency at different gate voltages.  Fig.~\ref{fig:fig4}(b) shows the Fourier transform of the oscillations in $1/B$ at each gate voltage. The oscillation frequency provides the area of the Fermi pocket.  One expects a linear relationship between the oscillation frequency and the total density: $B_F = (\phi_0/N) |n|$. Near the Dirac point at low densities, we observe a small oscillation frequency corresponding to the circular Fermi contour as shown in Fig.~\ref{fig:fig4}(c). As we increase the density, the slope gives $N=8$, as expected from the 2-fold layer, valley, and spin degeneracies. Near the insulating states, we find a single oscillation frequency with $N=4$. Calculated band structures present a Star-of-David Fermi contour, which suggests three possible electron orbits as illustrated in Fig.~\ref{fig:fig4}(e): (1) the outer star orbit, (2) the triangular orbits, and (3) the inner hexagon orbit. We overlaid the numerically extracted areas of these three types of orbits on top of the experimental data in Fig.~\ref{fig:fig4}(b), and only the triangular orbit fits with the experimental data. A similar scenario occurs for the Star-of-David Fermi contours around the valley points of the MBZ as shown in Fig.~\ref{fig:fig4}(d). This suggests that the crossing points of the two triangular orbits are protected \cite{shoenberg1984}. 

The large momentum mismatch between the original graphene $K$ and $K^\prime$ points provides a natural explanation for the suppressed hybridization. The MBZ arising from the moir\'{e} pattern folds the graphene band structures of both layers and creates degeneracies within it. The degree of hybridization at these degenerate crossings depends on the interlayer hopping amplitude: crucially, this amplitude varies exponentially with the momentum difference of the original states, with a characteristic momentum scale of the moir\'{e} wavevector $k_{\mathrm{SL}}$ \cite{shallcross2013,supplementary,bistritzer2011}. Since the $K^{(1)}$ and $K^{(2)}$ points (superscript labels the layer) are separated by a momentum less than $k_{\mathrm{SL}}$, the Dirac cones at $K^{(1,2)}$ hybridize strongly, and similarly for the $K^{\prime(1,2)}$ pair as well. These two pairs of hybridized Dirac cones form two time-reversed Fermi surfaces of opposite valley polarizations. Finally, while these two Fermi contours intersect within the MBZ, coupling these states requires a momentum difference corresponding to the inter-valley momentum of monolayer graphene [see Fig.~\ref{fig:fig4}(f)], which is much larger than $k_{\mathrm{SL}}$. The exponentially small interlayer hopping amplitude at this momentum leaves the crossings effectively unhybridized. As a result, we observe a single Fermi surface area consistent with the pair of triangular valley polarized orbits. 

In summary, we have experimentally studied the magnetotransport properties of high-quality TwBLG samples in the low twist angle regime, where we have observed insulating states induced by strong interlayer interactions. The larger than theoretically predicted gap sizes observed in the experiment indicate the possibility of other effects beyond the superlattice modulation and interlayer hybridization, such as strain and many-body interactions, therefore providing motivation for further theoretical and experimental studies in TwBLG.

\begin{acknowledgments}
We acknowledge helpful discussions with L. Fu, B. I. Halperin and B. Skinner. We also acknowledge fabrication assistance from Y. Bie. This work has been primarily supported by the National Science Foundation (DMR-1405221) for device fabrication, transport measurements, and data analysis (Y.C., J.Y.L., V.F., J.D.S-Y., P.J.H.), with additional support from the NSS Program, Singapore (J.Y.L.). This research has been funded in part by the Gordon and Betty Moore Foundation's EPiQS Initiative through Grant GBMF4541 to P.J.H.  This work made use of the Materials Research Science and Engineering Center Shared Experimental Facilities supported by the National Science Foundation (DMR-0819762) and of Harvard's Center for Nanoscale Systems, supported by the NSF (ECS-0335765). S.F. acknowledges support by the STC Center for Integrated Quantum Materials, NSF Grant No. DMR- 1231319, and E.K. acknowledges support by ARO MURI Award W911NF-14-0247.
\end{acknowledgments}


\begin{thebibliography}{99}

\bibitem{geim2013} A. K. Geim, I. V. Grigorieva. Nature. \textbf{499}, 419-425 (2013).
\bibitem{hunt2013} B. Hunt, J. D. Sanchez-Yamagishi, A. F. Young, M. Yankowitz, B. J. LeRoy, K. Watanabe, T. Taniguchi, P. Moon, M. Koshino, P. Jarillo-Herrero, R. C. Ashoori, Science \textbf{340}, 1427-1430 (2013).
\bibitem{woods2014} C. R. Woods, L. Britnell, A. Eckmann, R. S. Ma, J. C. Lu, H. M. Guo, X. Lin, G. L. Yu, Y. Cao, R. V. Gorbachev, \textit{et al.} Nature Phys. \textbf{10}, 451-456 (2014).
\bibitem{jung2015} J. Jung, A. M. DaSilva, A. H. MacDonald, S. Adam, Nature Comm. \textbf{6}, 6308 (2015).
\bibitem{sanjose2014} P. San-Jose, A. Guti\'{e}rrez-Rubio, M. Sturla, F. Guinea, Phys. Rev. B \textbf{90}, 075428 (2014).
\bibitem{song2013} J. C. W. Song, A. V. Shytov, L. S. Levitov, Phys. Rev. Lett. \textbf{111}, 266801 (2013).
\bibitem{santos2007} J. M. B. Lopes dos Santos, N. M. R. Peres, A. H. Castro Neto, Phys. Rev. Lett. \textbf{99}, 256802 (2007).
\bibitem{mele2010} E. J. Mele,  Phys. Rev. B \textbf{81}, 161405 (2010).
\bibitem{mele2011} E. J. Mele, Phys. Rev. B \textbf{84}, 235439 (2011).
\bibitem{bistritzer2011} R. Bistritzer, A. H. MacDonald, Proc. Natl. Acad. Sci. U. S. A. \textbf{108}(30), 12233-12237 (2011).
\bibitem{ohta2012} T. Ohta, J. T. Robinson, P. J. Feibelman, A. Bostwick, E. Rotenberg, T. E. Beechem, Phys. Rev. Lett. \textbf{109}, 186807 (2012).
\bibitem{havener2014} R. W. Havener, Y. Liang, L. Brown, L. Yang, J. Park, Nano Lett. \textbf{14}, 3353 (2014).
\bibitem{li2010} G. Li, A. Luican, J. M. B. Lopes dos Santos, A. H. Castro Neto, A. Reina, J. Kong, E. Y. Andrei, Nature Phys. \textbf{6}, 109 (2010).
\bibitem{luican2011} A. Luican, G. Li, A. Reina, J. Kong, R. R. Nair, K. S. Novoselov, A. K. Geim, E. Y. Andrei, Phys. Rev. Lett. \textbf{106}, 126802 (2011).
\bibitem{yan2012} W. Yan, M. Liu, R.-F. Dou, L. Meng, L. Feng, Z.-D. Chu, Y. Zhang, Z. Liu, J. C. Nie, L. He, Phys. Rev. Lett. \textbf{109}, 126801 (2012).
\bibitem{brihuega2012} I. Brihuega, P. Mallet, H. Gonz\'{a}lez-Herrero, G. Trambly de Laissardi\`{e}re, M. M. Ugeda, L. Maguad, J. M. G\'{o}mez-Rodr\'{i}guez, F. Yndur\'{a}in, J.-Y. Veuillen, Phys. Rev. Lett. \textbf{109}, 196802 (2012).
\bibitem{morell2010} E. Su\'{a}rez Morell, J. D. Correa, P. Vargas, M. Pacheco, Z. Barticevic, Phys. Rev. B \textbf{82}, 121407 (2010).
\bibitem{moon2012} P. Moon, M. Koshino, Phys. Rev. B \textbf{85}, 195458 (2012).
\bibitem{fang2016} S. Fang, E. Kaxiras. arXiv:1604.05371 (2016). 
\bibitem{wallbank2013} J. R. Wallbank, A. A. Patel, M. Mucha-Kruczynski, A. K. Geim, V. I. Fal'ko, Phys. Rev. B \textbf{87}, 245408 (2013).

\bibitem{lee2011} D. S. Lee, C. Riedl, T. Beringer, A. H. Castro Neto, K. von Klitzing, U. Starke, J. H. Smet, Phys. Rev. Lett. \textbf{107}, 216602 (2011).
\bibitem{schmidt2014} H. Schmidt, J. C. Rode, D. Smirnov, R. J. Haug, Nature Comm. \textbf{5}, 5742 (2014).
\bibitem{shallcross2013} S. Shallcross, S. Sharma, O. Pankratov, Phys. Rev. B \textbf{87}, 245403 (2013).

\bibitem{supplementary} See Supplemental Material.

\bibitem{wang2013} L. Wang, I. Meric, P. Y. Huang, Q. Gao, Y. Gao, H. Tran, T. Taniguchi, K. Watanabe, L. M. Campos, D. A. Muller, \textit{et al.} Science \textbf{342}, 614-617 (2013).

 \bibitem{kim2016} K. Kim, M. Yankowitz, B. Fallahazad, S. Kang, H. C. P. Movva, S. Huang, S. Larentiz, C. M. Corbet, T. Taniguchi, K. Watanabe, \textit{et al.} Nano Lett. \textbf{16}, 1989 (2016).
 
\bibitem{mott1968} N. F. Mott, J. Non-Cryst. Solids \textbf{1}, 1 (1968).

\bibitem{jerome1967} D. J\'{e}rome, T. M. Rice, W. Kohn, Phys. Rev. \textbf{158}, 462 (1967).
\bibitem{nandkishore2009} R. Nandkishore, L. Levitov, Phys. Rev. Lett. \textbf{104}, 156803 (2010).
\bibitem{bao2012} W. Bao, J. Velasco Jr., F. Zhang, L. Jing, B. Standley, D. Smirnov, M. Bockrath, A. H. MacDonald, C. N. Lau, Proc. Natl. Acad. Sci. U. S. A. \textbf{109}(27), 10802-10805 (2012).



\bibitem{degail2011} R. de Gail, M. O. Goerbig, F. Guinea, G. Montambaux, A. H. Castro Neto, Phys. Rev. B \textbf{84}, 045436 (2011).
\bibitem{jdsy2012} J. D. Sanchez-Yamagishi, T. Taychatanapat, K. Watanabe, T. Taniguchi, A. Yacoby, P. Jarillo-Herrero, Phys. Rev. Lett. \textbf{108}, 076601 (2012).

\bibitem{kim2016_2} During the preparation of this manuscript we became aware of Y. Kim, P. Herlinger, P. Moon, M. Koshino, T. Taniguchi, K. Watanabe, J. H. Smet,  arXiv:1605.05475 (2016).

\bibitem{fang2015} S. Fang, R. Kuate Defo, S. N. Shirodkar, S. Lieu, G. A. Tritsaris, E. Kaxiras, Phys. Rev. B \textbf{92}, 205108 (2015).

\bibitem{martin2003} G. W. Martin, D. L. Maslov, M. Y. Reizer, Phys. Rev. B \textbf{68}, 241309 (2003).

\bibitem{shoenberg1984} It could also be the case that any hybridization is so small that magnetic breakdown has already occurred, as outlined in D. Shoenberg, \textit{Magnetic Oscillations in Metals} (Cambridge University Press, Cambridge, 1984), Chap. 7.






\end{thebibliography}
\end{document}